\documentclass[twocolumn]{revtex4}
\usepackage[utf8]{inputenc}

\begin{document}

\title{Hydrodynamic model of skyrmions and vorticities in the spin-1 Bose-Einstein condensate in ferromagnetic phase at finite temperatures}

\author{Pavel A. Andreev}
\email{andreevpa@physics.msu.ru}
\affiliation{Faculty of physics, Lomonosov Moscow State University, Moscow, Russian Federation, 119991.}
\affiliation{Peoples Friendship University of Russia (RUDN University), 6 Miklukho-Maklaya Street, Moscow, 117198, Russian Federation}

\date{\today}

\begin{abstract}
Quantum hydrodynamic of skyrmions in spinor ultracold bosons at the finite temperature is described.
The limit regime of BEC appearing at the zero temperature is discussed either.
The skyrmions are found as the solution of the spin evolution equation for the ferromagnetic phase of spin-1 ultracold bosons.
It is demonstrated that the quantum part of the spin current gives major contribution in the formation of the skyrmions.
The interplay between the quantum (spin) vorticity and the classic vorticity is considered for the finite temperatures.
Bosons are presented as two fluids associated with the Bose-Einstein condensate and the normal fluid.
All effects are considered for the ferromagnetic phase.
\end{abstract}

%\pacs{}
% PACS, the Physics and Astronomy
                             % Classification Scheme.
%\keywords{spinor BEC, skyrmion, hydrodynamics, finite temperatures, quantum spin current.}
%Use showkeys class option if keyword

%03.75.Hh,
%03.75.Kk,
%67.85.Pq

\maketitle

%%%%%%%%%%TEXT

%\mbox{\boldmath $\Omega$}

\section{Introduction}

The skyrmions are among the well-known nonlinear phenomena \cite{Herbut PRL 06}, \cite{Tiurev NJP 18},
such as the solitons, shock waves, turbulence, spin turbulence and vortexes %vortices/vortexes
\cite{Tsatsos PR 16}, \cite{Fujimoto JLTP 14}, \cite{Stamper-Kurn RMP 13}.
All these phenomena are observed in ultracold gases,
but a possibility of skyrmions require nonzero spin of atoms.
The solitons can exist as the one dimensional structure
which is the area of increased or decreased concentration.
The skyrmion is the two dimensional structure of the spin vector field,
where the nontrivial distribution of the orientation of the spin vector is observed.
The skyrmions can be found in other mediums such as the chiral magnets
\cite{Muhlbauer Sc 09}, \cite{X.Z.Yu Nat Mat 11}, \cite{Heinze Nat P 11}, \cite{Everschor PRB 12},
\cite{Andreev EPL 20}, \cite{Trukhanova PTEP 20}, \cite{Yang}.

Skyrmions are considered as the elements for the information storage.
However, such complex structures are interesting from the fundamental point of view since they have nontrivial topological structure.
The skyrmions are characterized by the topological charge density and full topological charge,
which is related to the spin vorticity vector field.
The topological charge and the hydrodynamic helicity are possible invariants for the spin-1 ultracold bosons.
However, the interaction of the Bose-Einstein condensate with the normal fluid violates the conservation of this invariance.

Majority of papers on the ultracold bosons, boson-fermion and boson-boson mixtures are focused on the zero temperature regime
\cite{Nakamura PRA 14}, \cite{Fujimoto PRA 13}, \cite{Szirmai PRA 12}, \cite{Kudo PRA 10}, \cite{Andreev LP 19},
\cite{Deng PRA 20}, \cite{Saarikoski RMP 10}, \cite{Vinit PRA 17}, \cite{Andreev 1912}, \cite{Andreev LP 21}.
However, the number of work including the finite temperature effects in quantum gases
\cite{Mori JP B 18}, \cite{Roy PRA 14}, \cite{Gautam PRA 14}, \cite{Liu PRA 03}, \cite{Liu PRA 03 2},
\cite{Zaremba PRA 98}, \cite{Ticknor PRA 12}.
Particularly, the skyrmions in spin-1 BECs are considered at the zero temperature,
while the Zeeman term can be neglected either \cite{Herbut PRL 06}.
Here, we focused on the contribution of the excited states in the properties of the vorticity and the skyrmions.
Hence, both the tfinite temperature effects and the quantum fluctuations are discussed in the text below.

Vorticity is an important characteristic for the topological structures along with the topological charge density.
Hence, the spin related vorticity conservation is analysed via the quantum hydrodynamic equations.
The spin-1 BECs have nonpotential part of the velocity field.
Hence, the classical vorticity is nonzero.
The evolution of the spin vorticity happens together with the classic vorticity.
Each of them has the spin related sources,
but their superposition has no sources in the general vorticity evolution equation for the zero temperature BECs.
However, the finite temperature effects leads to the interaction related source even if the pressure of the bosons being in the excited states is dropped.

BEC of the spin-0 bosons is the single fluid,
while the finite temperature spin-0 bosons can be considered as two fluids: the BEC and the normal fluid.
The spin-1 BEC can be considered in two ways.
It can be described as the single fluid characterized by the spin density (and the nematic density) in addition to the concentration and the velocity field.
Otherwise, it can be considered as three fluids,
where each fluid is associated with the bosons of the fixed spin projection.
In this paper, we choose the single fluid picture of the spin-1 BEC.
Therefore, the finite temperature spin-1 bosons can be considered as two fluids similarly to the spin-0 notations.

This paper is organized as follows.
In Sec. II quantum hydrodynamic equations for the spin-1 ultracold finite temperature bosons being in the ferromagnetic phase are presented.
In Sec. III the evolution equations of the quantum (spin) and classic vorticities are obtained from the hydrodynamic equations
for the ferromagnetic phase of the spin-1 bosons.
In Sec. IV static equations for the structure of skyrmions are obtained from the hydrodynamic equations for different regimes.
In Sec. V a brief summary of obtained results is presented.

\section{Hydrodynamic equations for spin-1 BEC and normal fluid in the ferromagnetic phase}

Hydrodynamic theory of skyrmions in the spin-1 ultracold bosons at the finite temperatures is considered.
Therefore, we start this paper with the description of the basic hydrodynamic equations.
The finite temperature bosons with the majority of atoms being in the BEC state are under consideration.
We present bosons within the two fluid model,
where one fluid is associated with the BEC,
and the second fluid is associated with the normal fluid.
The normal fluid is the system of Bose atoms being in the quantum states with the energy above the minimal energy.
The quasi-classic point of view allows to consider a part of atoms in the BEC state and the rest of atoms being the normal fluid.
However, more accurate quantum picture suggests
that each particle has a probability to be in the quantum state with lowest energy and in the quantum states with higher energies.
The single microscopic many-particle wave function $\Psi_{S}(R,t)$ describes the dynamics of the whole system of bosons
separated on two subsystems at the macroscopic level of description.

Full concentration (the number density) of particles is
\begin{equation}\label{BECTSpin1Skyrmion concentration n definition}
n(\textbf{r},t)=\int dR\sum_{i=1}^{N}\delta(\textbf{r}-\textbf{r}_{i})\Psi_{S}^{\dagger}(R,t)\Psi_{S}(R,t), \end{equation}
where $^{\dagger}$ is the Hermitian conjugation.
The concentration $n(\textbf{r},t)$ is the scalar field in the three dimensional space of coordinates $\textbf{r}$,
which evolves in time $t$.
This is the definition of concentration via the $N$-particle microscopic wave function $\Psi_{S}(R,t)$
in the coordinate representation,
where $R$ is the vector in $3N$-dimensional configurational space: $R=\{\textbf{r}_{1}, .., \textbf{r}_{N}\}$,
and $dR$ is the element of volume in the configurational space.
Symbol $S$ presents the collection of $N$ spin indexes: $S=\{s_{1}, .., s_{N}\}$,
each index $s_{i}$ has three values $0, \pm1$ since we consider spin-1 bosons.
Concentration (\ref{BECTSpin1Skyrmion concentration n definition})
can be separated on two partial concentrations for the BEC and the normal fluid: $n=n_{B}+n_{n}$.
Definition (\ref{BECTSpin1Skyrmion concentration n definition}) is the first step of derivation of the hydrodynamic equations from the microscopic many-particle Schrodinger equation in the coordinate representation by the quantum hydrodynamic method suggested by
L. S. Kuz'menkov and S. G. Maksimov \cite{Maksimov QHM 99}, \cite{MaksimovTMP 2001}, \cite{Andreev PRA08}.
This method is applied to the spin-1 ultracold bosons \cite{Andreev 2102}.

We consider the ferromagnetic phase of spin-1 BECs,
while there are other phases such as polar phase, partially polarized BECs, antiferromagnetic phase
(which includes the nematic ordering of the system \cite{Symes PRA 17}).

Concentration (\ref{BECTSpin1Skyrmion concentration n definition}) is defined via the many-particle microscopic wave function $\Psi_{S}(R,t)$
in the coordinate representation.
The evolution of the wave function is governed by the Schrodinger equation.
Hence, the Schrodinger equation is used to derive the evolution equation for the concentration (\ref{BECTSpin1Skyrmion concentration n definition})
and other hydrodynamic functions.
Almost infinite set of hydrodynamic equations is truncated to include
the evolution of the concentration, the velocity field, spin vector field,
and the nematic tensor field
(the last one is reduced to the combination of spins for the ferromagnetic phase considered in this paper).
The truncation is made for the small temperature weakly interacting limit of systems of neutral atomic spin-1 bosons.
The finite temperature effects are presented via the normal fluid of bosons existing along the BEC.
It leads to the following set of hydrodynamic equations.
First, it is the continuity equations for the BEC and for the normal fluid \cite{Andreev 2102}
\begin{equation}\label{BECTSpin1Skyrmion continuity equation via v a}
\partial_{t}n_{a}+\nabla\cdot (n_{a}\textbf{v}_{a})=0, \end{equation}
where subindex $a$ stands for $B$ and $n$.
Vector $\textbf{v}_{a}$ is the hydrodynamic velocity field.
The velocity field is the ratio of the momentum density to the mass $m$ and concentration $n$.
The momentum density is defined via the many-particle wave function similarly to
the concentration (\ref{BECTSpin1Skyrmion concentration n definition}) \cite{Andreev 2102}.
The spinor BEC is characterized by the spin vector field
\begin{equation}\label{BECTSpin1Skyrmion S def} \textbf{s}(\textbf{r},t)=\frac{1}{n}
\int \Psi_{S}^{\dagger}(R,t)\sum_{i}\delta(\textbf{r}-\textbf{r}_{i})
(\hat{\textbf{F}}_{i}\Psi(R,t))_{S}dR, \end{equation}
which is the ratio of the spin density to the concentration $n$.
Definition (\ref{BECTSpin1Skyrmion S def}) contains
the single-particle spin matrixes
$$\hat{\textbf{F}}_{i}=\{\hat{F}_{x}, \hat{F}_{y}, \hat{F}_{z}\}$$
\begin{equation}\label{BECTSpin1Skyrmion spin matrixes} =\left\{\begin{array}{ccc}
                                                     \frac{1}{\sqrt{2}}\left(
                                                       \begin{array}{ccc}
                                                         0 & 1 & 0 \\
                                                         1 & 0 & 1 \\
                                                         0 & 1 & 0 \\
                                                       \end{array}
                                                     \right),
                                                      &
                                                      \frac{1}{\sqrt{2}}\left(
                                                       \begin{array}{ccc}
                                                         0 & -\imath & 0 \\
                                                         \imath & 0 & -\imath \\
                                                         0 & \imath & 0 \\
                                                       \end{array}
                                                     \right),
                                                     &
                                                     \left(
                                                       \begin{array}{ccc}
                                                         1 & 0 & 0 \\
                                                         0 & 0 & 0 \\
                                                         0 & 0 & -1 \\
                                                       \end{array}
                                                     \right)
                                                   \end{array}\right\}.
\end{equation}

The velocity field of the spin-0 BECs is the potential field,
so the vorticity, which is the rotation of the mass current, is equal to zero.
The curl of the velocity field for spin-1 BECs is nonzero.
Moreover, there is the generalized Mermin-Ho relation for the classic vorticity \cite{Yukawa PRA 12} %eq.s 29-31
$$\varepsilon^{\alpha\beta\gamma}\partial^{\beta} v^{\gamma}
=\frac{1}{2}s_{\mu}\varepsilon^{\alpha\beta\gamma}\partial^{\beta} (s^{\mu}v^{\gamma}+J_{Q}^{\mu\gamma})$$
\begin{equation}\label{BECTSpin1Skyrmion vorticity gen zero Yukawa}
+n_{\mu\nu}\varepsilon^{\alpha\beta\gamma}\partial^{\beta} (n^{\mu\nu}v^{\gamma}+J_{N,Q}^{\mu\nu\gamma}). \end{equation}
Here, the classic vorticity $\varepsilon^{\alpha\beta\gamma}\partial^{\beta} v^{\gamma}$
is expressed via the spin field and the nematic tensor field,
including their quantum fluxed.
The nematic tensor field $n^{\alpha\beta}$ has definition similar to the spin field (\ref{BECTSpin1Skyrmion S def})
at the corresponding replacement of operators $\hat{F}_{i}^{\alpha}$ to operators
$\hat{n}_{i}^{\alpha\beta}=(\hat{F}_{i}^{\alpha}\hat{F}_{i}^{\beta}+\hat{F}_{i}^{\beta}\hat{F}_{i}^{\alpha})/2$.
Equation
(\ref{BECTSpin1Skyrmion vorticity gen zero Yukawa})
can be rewritten as the superposition of similar structures
\begin{equation}\label{BECTSpin1Skyrmion sum of structures}
\sum_{i=0}^{2}c_{i}a^{\Gamma}\varepsilon^{\alpha\beta\gamma}\partial^{\beta} (a^{\Gamma}v^{\gamma}+A_{Q}^{\Gamma\gamma})=0, \end{equation}
where
$c_{0}=1$, $c_{1}=-1/2$, $c_{2}=-1$.
Each term in structure (\ref{BECTSpin1Skyrmion sum of structures}) is associated with different degrees of spin from $0$ to $2$,
where $2$ is the highest degree of spin matrixes for the spin-1 particles, which is not reduces to the smaller degree of spin matrixes.
For scalar $a^{\Gamma}$ reduces to 1,%=n/n
for the vector $a^{\Gamma}$ reduces to $S^{\mu}/n=s^{\mu}$,
for the second rank tensor $a^{\Gamma}$ presents to $N^{\mu\nu}/n=n^{\mu\nu}$,
$\Gamma$ is the set of corresponding number of tensor indexes.
The first term is the concentration related term.
It does not contain the quantum Bohm potential,
since the quantum Bohm potential is associated with the velocity field.
Expression $a^{\Gamma}v^{\gamma}+A_{Q}^{\Gamma\gamma}$ is the flux of function $a^{\Gamma}$,
where the flux includes the classic flux $a^{\Gamma}v^{\gamma}$
and the quantum part of flux $A_{Q}^{\Gamma\gamma}$,
but the thermal part is dropped due to the zero temperature nature of BECs.

Two Euler (the momentum balance) equations are
$$mn_{B}(\partial_{t}+(\textbf{v}_{B}\nabla))v^{\alpha}_{B}+\partial_{\beta}T^{\alpha\beta}_{B}$$
$$=-n_{B}\partial_{\alpha}V_{ext}
-g_{1}n_{B}\partial^{\alpha}n_{B}
-2g_{1}n_{B}\partial^{\alpha}n_{n}$$
\begin{equation}\label{BECTSpin1Skyrmion Euler for v B with T BaNF}
-g_{2}n_{B}\textbf{s}_{B}\partial^{\alpha}(n_{B}\textbf{s}_{B})
-2g_{2}n_{B}\textbf{s}_{B}\partial^{\alpha}(n_{n}\textbf{s}_{n}),
\end{equation}
and
$$mn_{n}(\partial_{t}+(\textbf{v}_{n}\nabla))v^{\alpha}_{n}+\partial_{\beta}T^{\alpha\beta}_{n}$$
$$=-n_{n}\partial_{\alpha}V_{ext}
-2g_{1}n_{n}\partial^{\alpha}n_{B}
-2g_{1}n_{n}\partial^{\alpha}n_{n}$$
\begin{equation}\label{BECTSpin1Skyrmion Euler for v NF with T BaNF}
-2g_{2}n_{n}\textbf{s}_{n}\partial^{\alpha}(n_{B}\textbf{s}_{B})
-2g_{2}n_{n}\textbf{s}_{n}\partial^{\alpha}(n_{n}\textbf{s}_{n}), \end{equation}
where
\begin{equation}\label{BECTSpin1Skyrmion g i def} g_{i}=\int U_{i}(r)d\textbf{r} \end{equation}
are the interaction constants.

The left-hand side of the Euler equation contains the kinematic effects including the quantum Bohm potential,
and the zero value of the kinetic pressure.
The right-hand side contains the action of the external field presented by the potential $V_{ext}$,
and the interparticle interaction.
The interaction is presented within two interaction constants $g_{i}$ (\ref{BECTSpin1Skyrmion g i def}).
Their definitions (\ref{BECTSpin1Skyrmion g i def}) contains the potentials (or their parts) of the interparticle interaction.
Function $U_{1}(r)$ is the spin independent part of the short-range interaction.
Function $U_{2}(r)$ is the part of the spin dependent potential of interaction
$\Delta \hat{H}=(1/2)\sum_{i,j\neq i}U_{2}(\mid r_{i}-r_{j}\mid)\hat{\textbf{F}}_{i}\cdot\hat{\textbf{F}}_{j}$.
The interaction terms in the Euler equations
(\ref{BECTSpin1Skyrmion Euler for v B with T BaNF}) and (\ref{BECTSpin1Skyrmion Euler for v NF with T BaNF})
(and in the spin evolution equation presented below)
are obtained in the first order by the interaction radius \cite{Andreev PRA08}.
The spin independent force field in equation (\ref{BECTSpin1Skyrmion Euler for v B with T BaNF})
$\textbf{F}_{B,1}=-g_{1}n_{B}\nabla(n_{B}+2n_{n})$
presents the action of the BEC and normal fluid on the bosons being in the BEC state.
The spin dependent part of the force field is presented via the spin densities $\textbf{S}_{a}=n_{a}\textbf{s}_{a}$.
The BEC-BEC interaction is presented by term with coefficient 1,
while the NF-BEC and NF-NF interactions contain the additional coefficient 2,
where NF stands for the normal fluid.
This distribution of coefficients is in agreement with spin-0 finite temperature bosons model presented in review \cite{Dalfovo RMP 99}
(see also \cite{Griffin PRB 96}).
%(see equations 127-129 on page 497).

The quantum Bohm potential consists of three parts \cite{Yukawa PRA 12}:
$$T_{a}^{\alpha\beta}=\frac{\hbar^{2}}{4m}\frac{1}{n}
\biggl(\partial^{\alpha}n\cdot \partial^{\beta}n
-n\partial^{\alpha}\partial^{\beta}n
\biggr)$$
$$+\frac{\hbar^{2}}{4m}n\biggl[
\frac{1}{2}[\partial^{\alpha}s^{\mu}\cdot\partial^{\beta}s^{\mu} -s^{\mu}\partial^{\alpha}\partial^{\beta}s^{\mu}] $$
\begin{equation}\label{BECTSpin1Skyrmion}
+[\partial^{\alpha}n^{\mu\nu}\cdot\partial^{\beta}n^{\mu\nu} -n^{\mu\nu}\partial^{\alpha}\partial^{\beta}n^{\mu\nu}]\biggr].
\end{equation}
The first group of terms consists purely of concentration $n$.
It coincides with the quantum Bohm potential for the spin-0 bosons.
Two other groups presents the quantum spin texture.
The second group of terms consists of the derivatives of the spin field $\textbf{s}$,
while the last group of terms consists of the derivatives of the nematic tensor field $n^{\alpha\beta}$
(the ratio of the nematic tensor density to the concentration).

The short-range interaction gives no contribution in the spin evolution equation in the first order by the interaction radius,
even its spin dependent part.
Hence, it can be written for two species in the following form
\begin{equation}\label{BECTSpin1Skyrmion spin evolution reduced a}
\partial_{t}s_{a}^{\alpha}+(\textbf{v}_{a}\cdot\nabla)s_{a}^{\alpha}+\frac{\partial_{\beta}J_{Q,a}^{\alpha\beta}}{n_{a}}=
\varepsilon^{\alpha z \gamma}\biggl(-\frac{p}{\hbar}s_{a}^{\gamma}
+2\frac{q}{\hbar}  n_{a}^{z\gamma}\biggr),\end{equation}
where $J_{Q,a}^{\alpha\beta}$ is the quantum part of the spin current
\begin{equation}\label{BECTSpin1Skyrmion spin current quantum}
J_{Q,a}^{\alpha\beta}=-\frac{\hbar}{m}n\varepsilon^{\alpha \gamma\delta}
\biggl(\frac{1}{4}s_{a}^{\gamma}\partial^{\beta} s_{a}^{\delta}
+n_{a}^{\gamma\mu}\partial^{\beta} n_{a}^{\delta\mu}\biggr). \end{equation}
The thermal part of the spin current existing for the normal fluid is neglected.

In accordance with equation (\ref{BECTSpin1Skyrmion spin evolution reduced a})
we can state
that there is no direct influence of the spin of BEC and spin of the normal fluid on each other.
However, this influence comes from the contribution of the velocity field in the second term and the concentration in the third term,
while these functions are bound for different species via the right-hand side of the Euler equations
(\ref{BECTSpin1Skyrmion Euler for v B with T BaNF}) and (\ref{BECTSpin1Skyrmion Euler for v NF with T BaNF}).

Complete hydrodynamic model of spin-ultracold bosons contains the nematic tensor evolution equation,
but the nematic tensor reduces to the spin field for the ferromagnetic phase
\begin{equation}\label{BECTSpin1Skyrmion} n^{\alpha\beta}_{a}=\frac{1}{2}(\delta^{\alpha\beta}+s^{\alpha}s^{\beta}). \end{equation}
Therefore, we do not present this equation here.
For the zero temperature regime it can be found in Ref. \cite{Yukawa PRA 12} (see equation 25).
For the finite temperature it is derived in Ref. \cite{Andreev 2102}.

%\begin{equation}\label{BECTSpin1Skyrmion}  \end{equation}

Hydrodynamic model (\ref{BECTSpin1Skyrmion continuity equation via v a})-(\ref{BECTSpin1Skyrmion spin evolution reduced a})
is equivalent to the set of non-linear Schrodinger equations
$$\imath\hbar\partial_{t}\hat{\Phi}_{B}=\Biggl(-\frac{\hbar^{2}\nabla^{2}}{2m}+V_{ext}
-p\hat{F}_{z}+q\hat{F}_{z}^{2}$$
\begin{equation}\label{BECTSpin1Skyrmion GP B}
+g_{1}(n_{B}+2n_{n})+g_{2}(\textbf{S}_{B}+2\textbf{S}_{n})\hat{\textbf{F}}\Biggr)\hat{\Phi}_{B}
,\end{equation}
and
$$\imath\hbar\partial_{t}\hat{\Phi}_{n}=\Biggl(-\frac{\hbar^{2}\nabla^{2}}{2m}+V_{ext}
-p\hat{F}_{z}+q\hat{F}_{z}^{2}$$
\begin{equation}\label{BECTSpin1Skyrmion GP like nf}
+2g_{1}(n_{B}+n_{n})+2g_{2}(\textbf{S}_{B}+\textbf{S}_{n})\hat{\textbf{F}}\Biggr)\hat{\Phi}_{n},\end{equation}
where
the three component vector macroscopic wave functions have the following structure
$\hat{\Phi}_{a}=\sqrt{n_{a}}e^{\imath m\phi_{a}/\hbar}\hat{z}_{a}$,
where $\phi_{a}$ is the part of the potential of the velocity field,
and $\hat{z}_{a}$ is the unit vector (the three component colomn).
The concentrations $n_{a}$ is the square of module of the macroscopic wave function
$n_{a}=\hat{\Phi}_{a}^{\dag}\hat{\Phi}_{a}$.
The spin densities $\textbf{S}_{a}$ are also represented via the macroscopic wave function
$\textbf{S}_{a}=\hat{\Phi}_{a}^{\dag}\hat{\textbf{F}}\hat{\Phi}_{a}$, $\textbf{S}_{a}=n_{a}\textbf{s}_{a}$.
Equations (\ref{BECTSpin1Skyrmion GP B}) and (\ref{BECTSpin1Skyrmion GP like nf})
are the finite temperature generalization of the spinor Gross-Pitaevskii equation
\cite{Stamper-Kurn RMP 13},
\cite{Kawaguchi PR 12},
\cite{Fujimoto PRA 14},
\cite{Oh PRL 14},
\cite{Ohmi JPSJ 98},
\cite{Ho PRL 98}.

\section{Topological charge, vorticity and hydrodynamic helicity}

The spin density leads to the existence to the topological scalar charge density defined as
\begin{equation}\label{BECTSpin1Skyrmion Topological charge}
q(\textbf{r})=\frac{1}{8\pi}
\varepsilon^{ij}\textbf{s}\cdot \partial_{i} \textbf{s} \times\partial_{j} \textbf{s}. \end{equation}
total topological charge
$Q=\int d\textbf{r} q(\textbf{r})$
is an invariant for two component BECs simultaneously described via the quasi-spinor two-component wave function \cite{Yang PRA 08}.
Hence, the mixture of BECs is described via the quasi-spin density
(see also \cite{Girvin PT 00}, \cite{Moon PRB 95}).
The topological charge density is closely related to the hydrodynamic spin vorticity
\begin{equation}\label{BECTSpin1Skyrmion vorticity spin def}
\Omega_{q,a}^{k}(\textbf{r},t)=\frac{\hbar}{m}\varepsilon^{ijk}\varepsilon^{\alpha\beta\gamma}
s_{a}^{\alpha}\cdot \partial_{i} s_{a}^{\beta} \partial_{j} s_{a}^{\gamma}. \end{equation}
The spin evolution equation allows to calculate the equation for evolution of the spin vorticity.
The classic hydrodynamic vorticity is equal to curl of velocity
$\Omega_{c}^{\alpha}=\varepsilon^{\alpha\beta\gamma}\partial^{\beta}v^{\gamma}$.
The topological charge can be connected to the spin vorticity $q=(8\pi m /\hbar)(\Omega_{q,a}^{x}+\Omega_{q,a}^{y}+\Omega_{q,a}^{z})$.

%\begin{equation}\label{BECTSpin1Skyrmion}  \end{equation}

%\begin{equation}\label{BECTSpin1Skyrmion}  ,\end{equation}

To start the analysis of the vorticity dynamical properties
we consider the evolution of the classic vorticities
$\mbox{\boldmath $\Omega$}_{a,c}=\nabla\times \textbf{v}_{a}$.
They are introduced for each species independently.
But their evolution follows from the corresponding Euler equations
(\ref{BECTSpin1Skyrmion Euler for v B with T BaNF})
and
(\ref{BECTSpin1Skyrmion Euler for v NF with T BaNF}).
The Euler equations are divided by corresponding concentration
and further taking curl of obtained equations.
It gives the equation of evolution of the classic vorticity for the spinor BEC and for the normal fluid
$$m\partial_{t}\Omega_{a,c}^{\alpha}-m[\nabla\times(\textbf{v}_{a}\times \mbox{\boldmath $\Omega$}_{a,c})]^{\alpha}$$
\begin{equation}\label{BECTSpin1Skyrmion vort c a 1}
+\varepsilon^{\alpha\beta\gamma}\partial^{\beta}\biggl(\frac{\partial^{\delta} T_{a}^{\gamma\delta}}{n_{a}}\biggr)
=-2g_{2}\varepsilon^{\alpha\beta\gamma}\partial^{\beta}s_{a}^{\delta}\cdot\partial^{\gamma}(n_{a'}s_{a'}^{\delta}),
 \end{equation}
where $a'$ stands for species different from $a$.
Equation (\ref{BECTSpin1Skyrmion vort c a 1}) shows
that the interaction between the BEC and normal fluid can be a source of the partial classic vorticities.
Moreover, the quantum Bohm potential $T_{a}^{\alpha\beta}$ requires further analysis.

The divergence of the first part of the quantum Bohm potential related to the concentration can be written in the following form
(see for instance \cite{Yukawa PRA 12} the third term in eq. 28)
\begin{equation}\label{BECTSpin1Skyrmion qBp div n part}
\partial_{\beta}T_{a,n}^{\alpha\beta}
=-\frac{\hbar^{2}}{2m}n_{a}
\partial^{\alpha}\frac{\triangle \sqrt{n_{a}}}{\sqrt{n_{a}}}. \end{equation}
Consequently,
$\varepsilon^{\alpha\beta\gamma}\partial^{\beta}(\partial_{\beta}T_{a,n}^{\alpha\beta}/n_{a})=0$.

The spin part of the quantum Bohm potential (for the ferromagnetic phase) can be transformed to
$$\partial_{\beta}T_{a,S}^{\alpha\beta}
=-\frac{\hbar^{2}}{8m}n_{a}\partial^{\alpha}(\partial^{\beta}s_{a}^{\mu}\cdot\partial^{\beta}s_{a}^{\mu})$$
\begin{equation}\label{BECTSpin1Skyrmion qBp div S part}
-\frac{\hbar^{2}}{4m}n_{a}s_{a}^{\mu}\partial^{\alpha}\biggl(\frac{\partial^{\beta}(n_{a}\partial^{\beta}s_{a}^{\mu})}{n_{a}}\biggr).
\end{equation}
Similar structure appears for the nematic part of the quantum Bohm potential:
$$\partial_{\beta}T_{a,N}^{\alpha\beta}
=-\frac{\hbar^{2}}{4m}n_{a}\partial^{\alpha}(\partial^{\beta}n_{a}^{\mu\nu}\cdot\partial^{\beta}n_{a}^{\mu\nu})$$
\begin{equation}\label{BECTSpin1Skyrmion qBp div N part}
-\frac{\hbar^{2}}{2m}n_{a}n_{a}^{\mu\nu}\partial^{\alpha}\biggl(\frac{\partial^{\beta}(n_{a}\partial^{\beta}n_{a}^{\mu\nu})}{n_{a}}\biggr).
\end{equation}

The curl of the first term of $\partial_{\beta}T_{a,n}^{\alpha\beta}/n_{a}$ for the spin and nematic parts gives the zero value,
while the second part of (\ref{BECTSpin1Skyrmion qBp div S part}) and (\ref{BECTSpin1Skyrmion qBp div N part})
create the source of vorticity.
The pressure of the BECs is equal to zero,
but the classical vorticity has the quantum source related to the spin density.

After the transformation of the quantum Bohm potential
equation (\ref{BECTSpin1Skyrmion vort c a 1}) transforms into:
$$m\partial_{t}\Omega_{a,c}^{\alpha}=m[\nabla\times(\textbf{v}_{a}\times \mbox{\boldmath $\Omega$}_{a,c})]^{\alpha}$$
$$+\varepsilon^{\alpha\beta\gamma}\partial^{\beta}s_{a}^{\mu}\cdot\partial^{\gamma} \biggl(\frac{\hbar^{2}}{4m}\frac{\partial^{\beta}(n_{a}\partial^{\beta}s_{a}^{\mu})}{n_{a}}\biggr)$$
$$+\varepsilon^{\alpha\beta\gamma}\partial^{\beta}n_{a}^{\mu\nu}\cdot\partial^{\gamma} \biggl(\frac{\hbar^{2}}{2m}\frac{\partial^{\beta}(n_{a}\partial^{\beta}n_{a}^{\mu\nu})}{n_{a}}\biggr)$$
\begin{equation}\label{BECTSpin1Skyrmion vort c a 2}
-2g_{2}\varepsilon^{\alpha\beta\gamma}\partial^{\beta}s_{a}^{\delta}\cdot\partial^{\gamma}(n_{a'}s_{a'}^{\delta}),
\end{equation}
where
$n_{a}^{\mu\nu}=(\delta^{\mu\nu}+s_{a}^{\mu}s_{a}^{\nu})/2$,
but we keep using nematic tensor to demonstrate structure of the obtained terms.

The classic vorticity evolution equation (\ref{BECTSpin1Skyrmion vort c a 2})
contains the following structures appearing from the quantum Bohm potential
\begin{equation}\label{BECTSpin1Skyrmion a mu def}
a_{a}^{\mu}=\frac{\hbar^{2}}{4m}\frac{\partial^{\beta}(n_{a}\partial^{\beta}s_{a}^{\mu})}{n_{a}}, \end{equation}
and
\begin{equation}\label{BECTSpin1Skyrmion a mu nu def}
a_{a}^{\mu\nu}=\frac{\hbar^{2}}{2m}\frac{\partial^{\beta}(n_{a}\partial^{\beta}n_{a}^{\mu\nu})}{n_{a}}. \end{equation}
These functions repeat themselves in the quantum part of the spin current
$$(\partial_{t}+\textbf{v}_{a}\cdot\nabla)s_{a}^{\alpha}$$
\begin{equation}\label{BECTSpin1Skyrmion spin evolution reduced via a mu and a mu nu}
-\frac{\varepsilon^{\alpha\beta\gamma}}{\hbar}[s_{a}^{\beta}a_{a}^{\gamma}+2n_{a}^{\beta\mu}a_{a}^{\gamma\mu}]
=\frac{\varepsilon^{\alpha z \gamma}}{\hbar}[-ps_{a}^{\gamma}+2q n_{a}^{z\gamma}], \end{equation}
which is demonstrated as a part of the spin evolution equation.

Moreover,
the contribution of the spin part of the quantum Bohm potential
$\varepsilon^{\alpha\beta\gamma}\partial^{\beta}s_{a}^{\mu}\cdot\partial^{\gamma}a_{a}^{\mu}$
and the nematic part of the quantum Bohm potential
$\varepsilon^{\alpha\beta\gamma}\partial^{\beta}n_{a}^{\mu\nu}\cdot\partial^{\gamma}a_{a}^{\mu\nu}$
in the evolution of the classic vorticity are equal to each other.
Hence, equation (\ref{BECTSpin1Skyrmion vort c a 2}) simplifies to
$$m\partial_{t}\Omega_{a,c}^{\alpha}=m[\nabla\times(\textbf{v}_{a}\times \mbox{\boldmath $\Omega$}_{a,c})]^{\alpha}$$
\begin{equation}\label{BECTSpin1Skyrmion vort c a 3}
+2\varepsilon^{\alpha\beta\gamma}\partial^{\beta}s_{a}^{\mu}\cdot\partial^{\gamma} a_{a}^{\mu}
-2g_{2}\varepsilon^{\alpha\beta\gamma}\partial^{\beta}s_{a}^{\delta}\cdot\partial^{\gamma}(n_{a'}s_{a'}^{\delta}),
\end{equation}

The classic vorticity is considered above.
However, there is vorticity associated with the spin field.
Its evolution can be calculated from the spin field evolution equation.

The spin vorticity (\ref{BECTSpin1Skyrmion vorticity spin def}) is defined via the spin field.
Hence, the spin evolution equation (\ref{BECTSpin1Skyrmion spin evolution reduced via a mu and a mu nu})
allows to derive the spin vorticity evolution equation
$$m\partial_{t}\Omega_{a,q}^{\alpha}
=m[\nabla\times(\textbf{v}_{a}\times \mbox{\boldmath $\Omega$}_{a,q})]^{\alpha}$$
\begin{equation}\label{BECTSpin1Skyrmion vort q a 1}
+2\varepsilon^{\alpha\beta\gamma}\partial^{\beta}s_{a}^{\mu}\cdot\partial^{\gamma} a_{a}^{\mu},
\end{equation}
where the contributions of the spin part of the quantum spin current
and the nematic part of the quantum spin current
are equal to each other.
They lead to the last term in equation (\ref{BECTSpin1Skyrmion vort q a 1}).

Full vorticity for species $a$ is the superposition of the classical and quantum vorticities.
Moreover, we choose the difference of these vorticities,
so there is no sources of the full vorticity in the absence of the second species:
\begin{equation}\label{BECTSpin1Skyrmion} \Omega_{a,f}^{\alpha}=\Omega_{a,c}^{\alpha}-\Omega_{a,q}^{\alpha}. \end{equation}
However, the normal fluid is considered along with the BEC:
$$m\partial_{t}\Omega_{a,f}^{\alpha}
=m[\nabla\times(\textbf{v}_{a}\times \mbox{\boldmath $\Omega$}_{a,f})]^{\alpha}$$
\begin{equation}\label{BECTSpin1Skyrmion vort f}
-2g_{2}\varepsilon^{\alpha\beta\gamma}\partial^{\beta}s_{a}^{\delta}\cdot\partial^{\gamma}(n_{a'}s_{a'}^{\delta}).
\end{equation}
Hence, there is the last term in equation (\ref{BECTSpin1Skyrmion vort f}),
which is the source of the vorticity.
Nevertheless, it is possible to consider the vorticity of all bosons as the
vorticities for the BEC $\Omega_{B,f}^{\alpha}$ and for the normal fluid $\Omega_{n,f}^{\alpha}$.
However, the spin-dependent part of the short-range interaction presented by the last term in equation (\ref{BECTSpin1Skyrmion vort f}) leads
to the nontrivial source of the combined vorticity.

%\begin{equation}\label{BECTSpin1Skyrmion}  \end{equation}

The hydrodynamic vorticity has the vector potential $\textbf{P}$:
$\mbox{\boldmath $\Omega$}=\nabla\times\textbf{P}$.
Therefore, it is possible to introduce the hydrodynamic helicity
$h=\int dV \textbf{P}\cdot\mbox{\boldmath $\Omega$}$.
It can be an invariant of motion of the medium.
For instance, if we consider the zero temperature BEC we obtain two term in equation (\ref{BECTSpin1Skyrmion vort f}).
In this case we have the conservation of the helicity $h$.
The nonconservation of the helicity can be related to the nonconservation of the topological charge.
Therefore, the skyrmions show an instability in this regime.

This result is obtained from the minimal coupling hydrodynamic model of finite temperature bosons,
where the thermal pressure and the thermal spin current are neglected for the normal fluid.
However, they exist as the small sources for the generalized hydrodynamic vorticity and helicity.
Similar contribution is discussed in Ref. \cite{Andreev PTEP 19}
for the charged spin-1/2 fermions.

\section{Skyrmions in spin-1 ultracold bosons}

\subsection{Skyrmions in the ferromagnetic phase at zero temperature, uniform limit}

First we consider skyrmions in the ferromagnetic phase at the zero temperature
for the uniform concentration.
Therefore, we repeat corresponding simplification of the set of hydrodynamic equations
(\ref{BECTSpin1Skyrmion continuity equation via v a}),
(\ref{BECTSpin1Skyrmion Euler for v B with T BaNF}),
(\ref{BECTSpin1Skyrmion Euler for v NF with T BaNF}),
(\ref{BECTSpin1Skyrmion spin evolution reduced a}),
\begin{equation}\label{BECTSpin1Skyrmion continuity equation via v B}
\partial_{t}n_{B}+\nabla\cdot (n_{B}\textbf{v}_{B})=0, \end{equation}
$$mn_{B}(\partial_{t}+(\textbf{v}_{B}\nabla))v^{\alpha}_{B}+\partial_{\beta}T^{\alpha\beta}_{B}=-n_{B}\partial_{\alpha}V_{ext}$$
\begin{equation}\label{BECTSpin1Skyrmion Euler for v B no T}
-g_{1}n_{B}\partial^{\alpha}n_{B}
-g_{2}n_{B}\textbf{s}_{B}\partial^{\alpha}(n_{B}\textbf{s}_{B}),
\end{equation}
and
\begin{equation}\label{BECTSpin1Skyrmion spin evolution reduced B}
(\partial_{t}+\textbf{v}_{B}\cdot\nabla)s_{B}^{\alpha}+\frac{\partial_{\beta}J_{Q}^{\alpha\beta}}{n_{B}}=
\varepsilon^{\alpha z \gamma}\biggl(-\frac{p}{\hbar}s_{B}^{\gamma}
+2\frac{q}{\hbar}  n_{B}^{z\gamma}\biggr).\end{equation}
This subsection includes description of the BEC only,
so subindex $B$ is mainly dropped.
The spin current is presented by equation (\ref{BECTSpin1Skyrmion spin current quantum}).
Another its representation can be found in equation (\ref{BECTSpin1Skyrmion spin evolution reduced via a mu and a mu nu}).
In the ferromagnetic phase the nematic tensor field $n^{\alpha\beta}$ reduces to the spin field
$n^{\alpha\beta}=(\delta^{\alpha\beta}+s^{\alpha}s^{\beta})/2$.

First consider the skyrmions for the uniform concentration $n=const$ (it corresponds to the zero external field)
and the macroscopically motionless medium $\textbf{v}=0$.
Equation (\ref{BECTSpin1Skyrmion continuity equation via v B})
is satisfied in this regime.

Skyrmions are the nonlinear structures of the magnetic moments or spins.
Therefore, the spin evolution equation is the major source of information for these structures.
The first group of terms in equation (\ref{BECTSpin1Skyrmion spin evolution reduced B}) goes to zero
$(\partial_{t}+\textbf{v}_{B}\cdot\nabla)s_{B}^{\alpha}=0$.
So, the major contribution follows from the quantum spin current $J_{Q}^{\alpha\beta}$.
The right-hand side also gives the nonzero contribution.

The spin field is the unit field for the ferromagnetic phase
\begin{equation}\label{BECTSpin1Skyrmion s unit vector}
\textbf{s}=\{\cos\Phi(\textbf{r}) \sin\Theta(\textbf{r}), \sin\Phi(\textbf{r}) \sin\Theta(\textbf{r}), \cos\Theta(\textbf{r})\}. \end{equation}
Functions $\Phi(\textbf{r})$ and $\Theta(\textbf{r})$ depend in general on the three dimensional coordinate $\textbf{r}$.
They are functions of time as well.
However, we are focused on the static regime,
where there is plane symmetry.
Choosing the $z$ axis perpendicular to the plane of symmetry
we find no dependence of functions $\Phi$ and $\Theta$ on coordinate $z$.
So, we have quasi-two dimensional geometry.
We choose the polar coordinates in $x-y$ plane.
Hence, we use $x=r\cos\varphi$ and $y=r\sin\varphi$.
Moreover, function $\Phi(\textbf{r})$ is reduced to $\pm\varphi$ for the skyrmion solution
for the Neel-type topological structures
\cite{Bera PRR 19}, \cite{Leonov NJP 16}.
To obtain the skyrmion solution function $\Theta(\textbf{r})$ can be considered as the function
of radial polar coordinate $\Theta(\textbf{r})=\Theta(r)$,
where $r=\sqrt{x^2+y^2}$.
Hence, equation (\ref{BECTSpin1Skyrmion s unit vector}) is reduced to
\begin{equation}\label{BECTSpin1Skyrmion s unit vector reduced}
\textbf{s}=\{\cos\varphi \sin\Theta(r), \sin\varphi \sin\Theta(r), \cos\Theta(r)\}. \end{equation}

Next, we substitute equation (\ref{BECTSpin1Skyrmion s unit vector reduced})
and assumptions described below equation (\ref{BECTSpin1Skyrmion s unit vector})
in the spin field evolution equation (\ref{BECTSpin1Skyrmion spin evolution reduced B})
and obtain nonlinear equation
which defines the structure of skyrmion
$$\frac{\hbar^{2}}{4m}\varepsilon^{\alpha\beta\gamma}s^{\beta}\triangle s^{\gamma}
+\frac{\hbar^{2}}{m}\varepsilon^{\alpha\beta\gamma}n^{\beta\mu}\triangle n^{\gamma\mu}$$
\begin{equation}\label{BECTSpin1Skyrmion spin evolution reduced B skyrmion}
=\varepsilon^{\alpha z \gamma}(p s^{\gamma}-2q n^{z\gamma}). \end{equation}

The left-hand side of equation (\ref{BECTSpin1Skyrmion spin evolution reduced B skyrmion}) is the divergence of the quantum spin current
$\partial_{\beta}J_{Q}^{\alpha\beta}$.
Moreover, the substitution of the nematic tensor field $n^{\alpha\beta}$ via the spin field $s^{\alpha}$ existing in the ferromagnetic phase
shows that the second term is equal to the first term in equation (\ref{BECTSpin1Skyrmion spin evolution reduced B skyrmion}).

Further analysis of equation (\ref{BECTSpin1Skyrmion spin evolution reduced B skyrmion})
gives the equation for function $\Theta(r)$.
The $z$ projection of equation (\ref{BECTSpin1Skyrmion spin evolution reduced B skyrmion})
is automatically satisfied.
The $x$ and $y$ projections of equation (\ref{BECTSpin1Skyrmion spin evolution reduced B skyrmion}) gives same result:
\begin{equation}\label{BECTSpin1Skyrmion skyrmion eq for Theta}
\Theta''+\frac{1}{r}\Theta'-\frac{\sin\Theta \cos\Theta}{r^{2}}=\frac{2m}{\hbar^{2}}\sin\Theta (p-q\cos\Theta), \end{equation}
where $\Theta'=d\Theta/dr$.
Coefficients $p$ and $q$ show
that the right-hand side (the left-hand side) of equation (\ref{BECTSpin1Skyrmion spin evolution reduced B skyrmion})
transforms into right-hand side (the left-hand side) of equation (\ref{BECTSpin1Skyrmion skyrmion eq for Theta}).
Equation (\ref{BECTSpin1Skyrmion skyrmion eq for Theta}) is different from equation 1 in Ref. \cite{Bera PRR 19}
since the right-hand side of equation presented in Ref. \cite{Bera PRR 19} contains one additional term.

The Euler equation (\ref{BECTSpin1Skyrmion Euler for v B no T}) includes the spin and nematic parts of the quantum Bohm potential,
which depends on the spin and its derivatives.
Hence, it can give nontrivial condition on the spin vector or the function $\Theta(r)$.
First, we represent the Euler equation (\ref{BECTSpin1Skyrmion Euler for v B no T})
for the ferromagnetic phase under additional condition of the zero velocity field:
\begin{equation}\label{BECTSpin1Skyrmion Euler for v B no T ferro v=0}
\partial_{\beta}T^{\alpha\beta}_{ferro}=-n_{B}\partial_{\alpha}V_{ext}-(g_{1}+g_{2})n_{B}\partial^{\alpha}n_{B},
\end{equation}
where
$$T_{ferro}^{\alpha\beta}=\frac{\hbar^{2}}{4m}\frac{1}{n}
\biggl(\partial^{\alpha}n\cdot \partial^{\beta}n
-n\partial^{\alpha}\partial^{\beta}n
\biggr)$$
\begin{equation}\label{BECTSpin1Skyrmion qBp ferro}
+\frac{\hbar^{2}}{2m}n\biggl[
\partial^{\alpha}s^{\mu}\cdot\partial^{\beta}s^{\mu}\biggr].
\end{equation}

Consider the uniform concentration and zero external field,
so equation (\ref{BECTSpin1Skyrmion Euler for v B no T ferro v=0}) simplifies to
\begin{equation}\label{BECTSpin1Skyrmion Euler for v B no T ferro v=0 uniform}
\partial_{\beta}[\partial^{\alpha}s^{\mu}\cdot\partial^{\beta}s^{\mu}]=0.
\end{equation}
So, the divergence of the spin dependent function is equal to zero.
Hence, the integration gives us the boundary condition at the infinity $(r^{\beta}/r)\partial^{\alpha}s^{\mu}\cdot\partial^{\beta}s^{\mu}=const$
for the normal component.

%\begin{equation}\label{BECTSpin1Skyrmion}  \end{equation}

%\begin{equation}\label{BECTSpin1Skyrmion}  \end{equation}

\subsection{Skyrmions in the ferromagnetic phase at zero temperature, nonuniform regime}

Let us make generalization of equations (\ref{BECTSpin1Skyrmion spin evolution reduced B skyrmion})
and (\ref{BECTSpin1Skyrmion skyrmion eq for Theta}) for the nonuniform medium $n\neq const$.
We assume that the concentration has the radial dependence only $n=n(r)$ in cylindrical coordinates.
Hence, we present the spin evolution equation (\ref{BECTSpin1Skyrmion spin evolution reduced B})
$$\frac{\hbar^{2}}{2m}\varepsilon^{\alpha\beta\gamma}
\biggl(s^{\beta}\triangle s^{\gamma}+\frac{1}{n}(\partial^{\nu}n)s^{\beta}\partial^{\nu}s^{\gamma}\biggr)$$
\begin{equation}\label{BECTSpin1Skyrmion spin evolution reduced B skyrmion nonuniform}
=\varepsilon^{\alpha z \gamma}[ps^{\gamma}-q\delta^{z\gamma}-qs^{z}s^{\gamma}], \end{equation}
where it is included that the nematic part of the quantum spin current is equal to the spin part of the quantum spin current for the ferromagnetic phase.
The spatial dependence of the concentration gives the additional contribution of the quantum spin current in compare with equation
(\ref{BECTSpin1Skyrmion spin evolution reduced B skyrmion}).
Here we use the substitution (\ref{BECTSpin1Skyrmion s unit vector reduced}) for the spin field to obtain generalization of equation
(\ref{BECTSpin1Skyrmion skyrmion eq for Theta})
\begin{equation}\label{BECTSpin1Skyrmion skyrmion eq for Theta nonuniform}
\Theta''+\frac{1}{r}\Theta'-\frac{\sin\Theta \cos\Theta}{r^{2}}+\frac{n'}{n}\Theta'
=\frac{2m}{\hbar^{2}}\sin\Theta (p-q\cos\Theta). \end{equation}
The concentration gives one additional term
which is placed as the last term on the left-hand side.

The Euler equation for the ferromagnetic phase spin-1 BECs with the zero velocity field is presented
by equation (\ref{BECTSpin1Skyrmion Euler for v B no T ferro v=0}).
For the zero external potential it can be represented
\begin{equation}\label{BECTSpin1Skyrmion Euler for v B no T ferro v=0 V=0 beta}
\partial_{\beta}T^{\alpha\beta}_{ferro}=-\frac{1}{2}(g_{1}+g_{2})\delta^{\alpha\beta}\partial^{\beta}n_{B}^{2}.
\end{equation}
Hence, it can be also integrated
\begin{equation}\label{BECTSpin1Skyrmion Euler for v B no T ferro v=0 V=0 beta integrated}
(r^{\beta}/r)(T^{\alpha\beta}_{ferro}+\frac{1}{2}(g_{1}+g_{2})\delta^{\alpha\beta}n_{B}^{2})=const.
\end{equation}
Therefore, equation (\ref{BECTSpin1Skyrmion Euler for v B no T ferro v=0 V=0 beta})
and condition (\ref{BECTSpin1Skyrmion Euler for v B no T ferro v=0 V=0 beta integrated})
show interplay the derivatives of concentration and interaction on the distribution of spin field.

%\begin{equation}\label{BECTSpin1Skyrmion}  \end{equation}

%\begin{equation}\label{BECTSpin1Skyrmion}  \end{equation}

%\begin{equation}\label{BECTSpin1Skyrmion}  \end{equation}

\subsection{Skyrmions for the full spin polarization at nonzero temperature: contribution of the normal fluid}

The spin evolution of BEC has no direct influence of the normal fluid since no contribution of the normal fluid
in equation (\ref{BECTSpin1Skyrmion spin evolution reduced a}).
However, the Euler equation can give contribution in the spin field distribution
via the boundary condition (\ref{BECTSpin1Skyrmion Euler for v B no T ferro v=0 V=0 beta integrated}).
Consideration of the nonzero temperatures leads to the contribution of the normal fluid in the Euler equation for the BEC.
It appears via the additional force field
\begin{equation}\label{BECTSpin1Skyrmion F acting on BEC}
F^{\alpha}=-2g_{1}n_{B}\partial^{\alpha}n_{n}-2g_{2}n_{B}\textbf{s}_{B}\partial^{\alpha}(n_{n}\textbf{s}_{n}). \end{equation}
It simplifies for the uniform concentrations of BEC and normal fluid:
\begin{equation}\label{BECTSpin1Skyrmion F acting on BEC uniform n}
F^{\alpha}=-2g_{2}n_{0B}n_{0n}\textbf{s}_{B}\partial^{\alpha}\textbf{s}_{n}. \end{equation}
We assume that the small temperature normal fluid is in the ferromagnetic state.
Hence, vector $\textbf{s}_{n}$ has structure presented by equation (\ref{BECTSpin1Skyrmion s unit vector reduced}).
Substitution of vectors $\textbf{s}_{B}$ and $\textbf{s}_{n}$ in equation (\ref{BECTSpin1Skyrmion F acting on BEC uniform n})
gives the zero value of the force field.
Hence, there is nonzero force field contribution for the nonuniform bosons only
\begin{equation}\label{BECTSpin1Skyrmion} F^{\alpha}=-2(g_{1}+g_{2}\cos(\Theta_{B}-\Theta_{n}))n_{B}\partial^{\alpha}n_{n}, \end{equation}
where the difference of angular functions $\Theta_{B}$ and $\Theta_{n}$ modulate the strength of interaction.

%\begin{equation}\label{BECTSpin1Skyrmion}  \end{equation}

\subsection{Skyrmions for the full spin polarization at the zero temperature: contribution of the quantum fluctuations}

The nonzero temperature provides the contribution of the bosons being in the excited quantum states.
However, the evolution of the BEC itself leads to the appearance of particles in the excited states.
It is called the quantum fluctuations
which are caused by the interaction of bosons.
The simple model suggests that
the quantum fluctuations can be included in the non-linear Schrodinger equation as the additional interaction term proportional to the fourth order
of the macroscopic wave function \cite{Bisset PRA 16}, \cite{Examilioti JP B 20}, \cite{Aybar PRA 19}:
\begin{equation} \label{BECTSpin1Skyrmion GP with LHY}
\imath\hbar\partial_{t}\Phi
=-\frac{\hbar^{2}}{2m}\triangle\Phi+(V_{ext} +g_{1}\mid\Phi\mid^{2}
+g_{qf}\mid\Phi\mid^{3})\Phi, \end{equation}
where
\begin{equation} \label{BECTSpin1Skyrmion g qf}
g_{qf}=\frac{32}{3\sqrt{\pi}}g_{1}\sqrt{a_{1}^{3}},\end{equation}
with the standard relation between the s-scattering amplitude and the interaction constant $g=4\pi\hbar^{2}a/m$.
Its appearance is associated with the Lee-Huang-Yang energy.
Equation (\ref{BECTSpin1Skyrmion GP with LHY}) is obtained for the spin-0 BECs,
but it shows the form of contribution of the quantum fluctuations.
Equation (\ref{BECTSpin1Skyrmion GP with LHY}) allows to introduce the effective interaction constant,
which depends on the distribution of bosons:
$\tilde{g}_{1}=g_{1}+g_{qf}\mid\Phi\mid$.
Similar generalization of the second interaction constant describing the spin dependent part of the interaction
can be suggested $\tilde{g}_{2}=g_{2}+g_{qf,2}\mid\Phi\mid$.
It leads to the zero temperature nonlinear Pauli equation for the spin-1 BECs:
$$\imath\hbar\partial_{t}\hat{\Phi}_{B}=\Biggl(-\frac{\hbar^{2}\nabla^{2}}{2m}+V_{ext}
-p\hat{F}_{z}+q\hat{F}_{z}^{2}+g_{1}\mid\hat{\Phi}_{B}\mid^{2}$$
\begin{equation} \label{BECTSpin1Skyrmion GP with LHY spin}
+g_{2}\mid\hat{\Phi}_{B}\mid^{2}\textbf{s}_{B}\hat{\textbf{F}}
+g_{qf}\mid\hat{\Phi}\mid^{3}+g_{qf,2}\mid\hat{\Phi}\mid^{3}\textbf{s}_{B}\hat{\textbf{F}})\Biggr)\hat{\Phi}_{B}, \end{equation}
where
\begin{equation} \label{BECTSpin1Skyrmion g qf}
g_{qf,2}=\frac{32}{3\sqrt{\pi}}g_{2}\sqrt{a_{2}^{3}},\end{equation}
and $g_{2}=4\pi\hbar^{2}a_{2}/m$.

If we consider the quantum fluctuations via the hydrodynamic notions
we see that
the presence of the particles in the excited states is related to the nonzero kinetic pressure.
Hence, it is necessary to consider the pressure in more details to get an equation of state
similar to the Lee-Huang-Yang expression (\ref{BECTSpin1Skyrmion GP with LHY}).
Corresponding derivation is made by the quantum hydrodynamic method.
After the derivation of the Euler equation
we have the pressure contribution $p^{\alpha\beta}_{B}$.
If we include that
there are particles in the excited states
due to the BEC dynamics
we cannot drop the pressure.
Hence, we have additional term on the left-hand side of the Euler equation:
$$mn_{B}(\partial_{t}+(\textbf{v}_{B}\nabla))v^{\alpha}_{B}+\partial_{\beta}T^{\alpha\beta}_{B}+\partial_{\beta}p^{\alpha\beta}_{B,qf}$$
\begin{equation}\label{BECTSpin1Skyrmion Euler for v B no T with p}
=-n_{B}\partial_{\alpha}V_{ext}
-g_{1}n_{B}\partial^{\alpha}n_{B}
-g_{2}n_{B}\textbf{s}_{B}\partial^{\alpha}(n_{B}\textbf{s}_{B}).
\end{equation}
To understand the contribution of pressure we need to give derivation of the pressure evolution equation.
We do not present here the derivation,
but some details can be found in Ref. \cite{Andreev 2005} (including the supplementary materials).
More technical details can be found in Refs. \cite{Andreev 2101} and \cite{Andreev 2001},
while these papers are on different subjects.
The simplified forms of the required hydrodynamic equations are presented below.

Quasi-linear local density approximation for the pressure evolution equation:
\begin{equation}\label{BECTSpin1Skyrmion p evol simp}
\partial_{t}p^{\alpha\beta}_{B,qf}+\partial_{\gamma}Q^{\alpha\beta\gamma}_{B,qf}=0 \end{equation}
while the general form can be found in Refs. \cite{Andreev 2005}, \cite{Andreev 2101}, \cite{Andreev 2007}, \cite{Andreev 2009}.

The simplified pressure evolution equation (\ref{BECTSpin1Skyrmion p evol simp}) presents the following information existing in its general form.
The pressure evolution equation does not contain any interaction
(at least in the first order by the interaction radius).
Hence, the nonzero pressure is caused by the pressure flux $Q^{\alpha\beta\gamma}_{B,qf}$,
so equation (\ref{BECTSpin1Skyrmion p evol simp}) is a continuity equation for the pressure.
If we neglect the quantum fluctuations both functions $p^{\alpha\beta}_{B,qf}$ and $Q^{\alpha\beta\gamma}_{B,qf}$ are equal to zero.
It is known that
the quantum fluctuations are caused by the interaction.
Therefore, it is necessary to derive equation for the evolution of the pressure flux $Q^{\alpha\beta\gamma}_{B,qf}$ to obtain the contribution of the interaction in the pressure dynamics.

Here, we drop the kinematic terms in the pressure flux evolution equation to show the appearance of tensor $Q^{\alpha\beta\gamma}_{B,qf}$
from the interaction:
\begin{equation}\label{BECTSpin1Skyrmion Q evol simpl with spin} \partial_{t}Q^{\alpha\beta\gamma}_{B,qf}=\frac{\hbar^{2}}{4m^{3}}
I_{0}^{\alpha\beta\gamma\delta} (g_{3}n_{B}\partial^{\delta}n_{B}
+g_{4}n_{B}\textbf{s}_{B}\partial^{\delta}(n_{B}\textbf{s}_{B})),\end{equation}
where
\begin{equation} \label{BECTSpin1Skyrmion def g 4} g_{i+2}=\frac{2}{3}\int d\textbf{r} U_{i}''(r). \end{equation}
The ferromagnetic phase leads to the simplification of the interaction similar to the simplification of the Euler equation:
\begin{equation}\label{BECTSpin1Skyrmion Q evol simpl ferro} \partial_{t}Q^{\alpha\beta\gamma}_{B,qf}
=\frac{\hbar^{2}}{4m^{3}}
I_{0}^{\alpha\beta\gamma\delta} (g_{3}+g_{4})n\partial^{\delta}n.
\end{equation}
Equation (\ref{BECTSpin1Skyrmion Q evol simpl ferro}) shows that the spin field contribution in the interaction reduces to the concentration for the ferromagnetic phase of spin-1 BECs.
It is similar to the properties of the force field in the Euler equation.

The structure of the general form of the pressure evolution equation and the third rank tensor evolution are highly nonlinear.
The simplified pressure flux evolution equation is nonlinear either.
The simplification of the quantum fluctuation discussed in Ref. \cite{Andreev 2007} is not suitable for skyrmions
since it is based on the linear approach on the small amplitude perturbation.
The nonlinear Schrodinger equation with the fourth order nonlinearity is based on the local density approximation,
since it is based on the expression obtained for the uniform medium.
All described approaches shows that
the quantum fluctuations gives no contribution in the skyrmions at the uniform concentration.

The hydrodynamic model of the quantum fluctuations presented by equations
(\ref{BECTSpin1Skyrmion Euler for v B no T with p})-(\ref{BECTSpin1Skyrmion Q evol simpl ferro})
shows the direction for the generalization of the hydrodynamic equations in order to find full picture of hydrodynamic quantum fluctuations
and their role in the vortical dynamic affecting the skyrmion structure.
The quantum fluctuations are related to the presence of particles in the excited states.
It leads to the additional spin current along with the pressure dynamics discussed above.
Existence of the spin current caused by the quantum fluctuations can give direct contribution in the spin evolution equation
(\ref{BECTSpin1Skyrmion spin evolution reduced B skyrmion})
or (\ref{BECTSpin1Skyrmion skyrmion eq for Theta}).
Therefore, it can give some modification of the skyrmions structure or its stability.

%\begin{equation}\label{BECTSpin1Skyrmion}  \end{equation}

\section{Conclusion}

Hydrodynamic model of skyrmions has been presented.
The skyrmions have been considered in the static regime of BECs.
The ferromagnetic phase of spin-1 BECs and the mixture of the BECs and the normal fluids have been considered.
In this regime the terms containing the nematic tensor are reduced to the spin field.
Moreover, the nematic parts of the quantum Bohm potential and the quantum spin current is equal to their spin parts.
It has been found that
the quantum part of the spin current gives major contribution in the formation of the skyrmions in BECs.
It appears from the part of the spin current existing at the uniform concentration
while the spin field orientation changes in space.
The Zeeman terms gives the contribution in the equation for skyrmions as well.

The skyrmions are characterized by the topological charge and the topological charge density
which is related with the spin part of the vorticity.
The spin vorticity has been considered together with the classic hydrodynamic vorticity associated with the velocity evolution.
Evolution of the spin vorticity has been derived from the hydrodynamic equations.
In this case, the source of the spin vorticity related
to the spin current of the quantum spin current disappears.
It is canceled by the similar term in the classic vorticity equation
which appears from the spin part of the quantum Bohm potential.
However, the term caused by the spin dependent part of the short-range interaction gives
the additional source of the classic vorticity for the nonzero temperature (presented by the normal fluid).
Hence, the full vorticity has nontrivial souses for the finite temperatures.
But no sources exist at the zero temperature.
Therefore, the hydrodynamic helicity is conserved in the zero temperature limit.
Furthermore, the influence of the quantum fluctuations in the properties of the vorticity is discussed for the zero temperature limit.
As a part of discussion for the further generalizations of hydrodynamic model for the spin-1 BECs.

\section{Acknowledgements}

Work is supported by the Russian Foundation for Basic Research (grant no. 20-02-00476).
This paper has been supported by the RUDN University Strategic Academic Leadership Program.

%\cite{}

\end{document}